\documentclass[12pt]{article}
\usepackage{amsmath,amssymb,color}
\usepackage[dvips]{graphicx}
\usepackage[normalem]{ulem} 
\usepackage{cancel}
\def\red#1{{\color{red} #1}}
\def\affil#1{{\normalsize #1}}  
\begin{document}

\def\prg#1{\medskip\noindent{\bf #1}}  \def\ra{\rightarrow}
\def\lra{\leftrightarrow}              \def\Ra{\Rightarrow}
\def\nin{\noindent}                    \def\pd{\partial}
\def\dis{\displaystyle}                \def\inn{\hook}
\def\grl{{GR$_\Lambda$}}               \def\Lra{{\Leftrightarrow}}
\def\cs{{\scriptstyle\rm CS}}          \def\ads3{{\rm AdS$_3$}}
\def\Leff{\hbox{$\mit\L_{\hspace{.6pt}\rm eff}\,$}}
\def\bull{\raise.25ex\hbox{\vrule height.8ex width.8ex}}
\def\ric{{Ric}}                        \def\tric{{(\widetilde{Ric})}}
\def\tmgl{\hbox{TMG$_\Lambda$}}
\def\Lie{{\cal L}\hspace{-.7em}\raise.25ex\hbox{--}\hspace{.2em}}
\def\sS{\hspace{2pt}S\hspace{-0.83em}\diagup}   \def\hd{{^\star}}
\def\dis{\displaystyle}                \def\ul#1{\underline{#1}}
\def\mb#1{\hbox{{\boldmath $#1$}}}     \def\grp{{GR$_\parallel$}}
\def\irr#1{^{(#1)}}                    \def\ub#1{\underbrace{#1}}

\def\hook{\hbox{\vrule height0pt width4pt depth0.3pt
  \vrule height7pt width0.3pt depth0.3pt
  \vrule height0pt width2pt depth0pt}\hspace{0.8pt}}
\def\semidirect{\;{\rlap{$\supset$}\times}\;}
\def\first{\rm (1ST)}       \def\second{\hspace{-1cm}\rm (2ND)}
\def\bm#1{\hbox{{\boldmath $#1$}}}
\def\nb#1{\marginpar{{\large\bf #1}}}
\def\ir#1{{}^{(#1)}}  \def\ArcTan{\text{ArcTan}}

\def\G{\Gamma}        \def\S{\Sigma}        \def\L{{\mit\Lambda}}
\def\D{\Delta}        \def\Th{\Theta}
\def\a{\alpha}        \def\b{\beta}         \def\g{\gamma}
\def\d{\delta}        \def\m{\mu}           \def\n{\nu}
\def\th{\theta}       \def\k{\kappa}        \def\l{\lambda}
\def\vphi{\varphi}    \def\ve{\varepsilon}  \def\p{\pi}
\def\r{\rho}          \def\Om{\Omega}       \def\om{\omega}
\def\s{\sigma}        \def\t{\tau}          \def\eps{\epsilon}
\def\nab{\nabla}      \def\btz{{\rm BTZ}}   \def\heps{\hat\eps}
\def\bt{{\bar t}}     \def\br{{\bar r}}     \def\bth{{\bar\theta}}
\def\bvphi{{\bar\vphi}}
\def\aa{{\bar a}}     \def\bb{{\bar b}}
\def\bx{{\bar x}}     \def\by{{\bar y}}     \def\bom{{\bar\om}}
\def\tphi{{\tilde\vphi}}  \def\tt{{\tilde t}}

\def\tG{{\tilde G}}   \def\cF{{\cal F}}      \def\bH{{\bar H}}
\def\cL{{\cal L}}     \def\cM{{\cal M }}     \def\cE{{\cal E}}
\def\cH{{\cal H}}     \def\hcH{\hat{\cH}}
\def\cK{{\cal K}}     \def\hcK{\hat{\cK}}    \def\cT{{\cal T}}
\def\cO{{\cal O}}     \def\hcO{\hat{\cal O}} \def\cV{{\cal V}}
\def\tom{{\tilde\omega}}                     \def\cE{{\cal E}}
\def\cR{{\cal R}}    \def\hR{{\hat R}{}}     \def\hL{{\hat\L}}
\def\tb{{\tilde b}}  \def\tA{{\tilde A}}     \def\tv{{\tilde v}}
\def\tT{{\tilde T}}  \def\tR{{\tilde R}}     \def\tcL{{\tilde\cL}}
\def\hy{{\hat y}\hspace{1pt}}  \def\tcO{{\tilde\cO}}
\def\bA{{\bar A}}     \def\bB{{\bar B}}      \def\bC{{\bar C}}
\def\bG{{\bar G}}     \def\bD{{\bar D}}      \def\bH{{\bar H}}
\def\bK{{\bar K}}     \def\bL{{\bar L}}      \def\bE{{\bar E}}
\def\pgp{\text{PG$^{+}$}}             \def\pgm{\text{PG$^{-}$}}
\def\kadsp{\text{Kerr-AdS{}$^{+}$}}   \def\kads{\text{Kerr-AdS}}
\def\kadsm{\text{Kerr-AdS$^{-}$}}   \def\bS{{\bar S}}
\def\km{\text{Kerr$^{-}$}}
\def\chm{\checkmark}  \def\chmr{\red{\chm}}
\def\nn{\nonumber}                    \def\vsm{\vspace{-7pt}}
\def\be{\begin{equation}}             \def\ee{\end{equation}}
\def\ba#1{\begin{array}{#1}}          \def\ea{\end{array}}
\def\bea{\begin{eqnarray} }           \def\eea{\end{eqnarray} }
\def\beann{\begin{eqnarray*} }        \def\eeann{\end{eqnarray*} }
\def\beal{\begin{eqalign}}            \def\eeal{\end{eqalign}}
\def\lab#1{\label{eq:#1}}             \def\eq#1{(\ref{eq:#1})}
\def\bsubeq{\begin{subequations}}     \def\esubeq{\end{subequations}}
\def\bitem{\begin{itemize}}           \def\eitem{\end{itemize}}
\renewcommand{\theequation}{\thesection.\arabic{equation}}

\title{Thermodynamics of Kerr-AdS black holes\\ in general Poincar\'e gauge theory}

\author{M. Blagojevi\'c and B. Cvetkovi\'c\footnote{
        Email addresses: \texttt{mb@ipb.ac.rs, cbranislav@ipb.ac.rs}} \\
\affil{Institute of Physics, University of Belgrade,
                      Pregrevica 118, 11080 Belgrade, Serbia} }
\date{}                                                 
\maketitle

\begin{abstract}
A Hamiltonian variational approach is used to study asymptotic charges and entropy of Kerr-AdS black holes in the general Poincar\'e gauge theory, with both even and odd parity modes. The results turn out to be the same as those found earlier in the sector of parity invariant Lagrangians.
\end{abstract}

\section{Introduction}
\setcounter{equation}{0}

The Poincar\'e gauge theory (PG) is a well-founded approach to gravity based on gauging the Poincar\'e group of spacetime symmetries \cite{x1}. The dynamical content of PG is expressed in terms of a Riemann-Cartan (RC) geometry of spacetime, characterized by the presence of two field strengths, the torsion $T^i$ and the curvature $R^{ij}$; for a comprehensive analysis of the subject, see the reader \cite{x2}, the lectures \cite{x3} and/or the monographs \cite{x4}.

In the last half a century, various dynamical and geometric aspects of PG, as well as its relation to physics, have been intensively studied. Thus, successes in constructing a number of black hole solutions \cite{x2} inspired a detailed analysis of their \emph{asymptotic charges}, energy and angular momentum; for an advanced exposition of the subject, see Ref. \cite{x5}. However, it is somewhat surprising that systematic studies of \emph{black hole entropy} were rather neglected in the literature. Quite recently, the subject came to life in the work \cite{x6}, where the idea that ``black hole entropy is the Noether charge" \cite{x7} was given a natural \emph{Hamiltonian extension}.
In the papers  \cite{x8,x9,x10}, this approach was used to study both the asymptotic charges and entropy of stationary black holes in PG with  \emph{parity invariant} Lagrangians. The results obtained for spherically and axially symmetric solutions on the Minkowski or anti-de Sitter (AdS) background confirmed the validity of the first law of black hole thermodynamics. On the other hand, in the last decade, one can notice an increased interest in exploring various dynamical aspects of the general \emph{parity violating} PG, such as cosmological applications, exact solutions and particle spectrum \cite{x11,x12,x13,x14,x15,x16,x17}. In the present paper, our attention is focused on exploring energy, angular momentum and entropy of Kerr-AdS black holes in the general PG, with both even and odd parity modes.

The paper is organized as follows. In Sec. \ref{sec2}, we give a short account  of the general PG \cite{x2,x17}, as well as an outline of the variational Hamiltonian approach to the thermodynamic charges of black holes with torsion \cite{x6}. In Sec. \ref{sec3}, we analyze geometric and dynamical aspects of \kads\ black holes, as a preparation for studying their thermodynamic charges. Then, in sections \ref{sec4} and \ref{sec5}, we apply the Hamiltonian approach to calculate the \kads\ asymptotic charges and entropy. The results are summarized in Sec. \ref{sec6}, and the Appendix is devoted to some technical details.

Our general conventions remain the same as in Refs. \cite{x9,x10}. The Latin indices $(i,j,\dots)$ refer to the local Lorentz frame, the Greek indices $(\m,\n,\dots)$ refer to the coordinate frame, $b^i$ is the orthonormal coframe (tetrad) dual to the frame $h_i$, with $h_i\inn b^k= \d_i^k$, the local Lorentz
metric is $\eta_{ij}=(1,-1,-1,-1)$, and $\om^{ij}$ is a metric compatible (antisymmetric) connection. The exterior product of forms is implicit, the volume 4-form is $\heps= b^0b^1b^2b^3$, the Hodge dual of a form $\a$ is $\hd\a$, $\hd 1 = \heps$, and the totally antisymmetric symbol is
normalized to $\ve_{0123}=+1$.

\section{PG dynamics and boundary terms}\label{sec2}
\setcounter{equation}{0}

Basic dynamical variables of PG are the tetrad field $b^i$ and the antisymmetric spin connection $\om^{ij}$ (1-forms), the gauge potentials associated to the translation and Lorentz subgroups of the Poincar\'e group, respectively. The corresponding field strengths are the torsion $T^i=d b^i+\om^i{}_k b^k$ and the curvature $R^{ij}=d\om^{ij}+\om^i{}_k\om^{kj}$, and the underlying structure of spacetime is characterized by a Riemann-Cartan geometry. In the absence of matter, dynamical properties of PG are determined by the gravitational Lagrangian $L_G(b^i,T^i,R^{ij})$ (4-form), which is assumed to be \emph{at most quadratic} in the field strengths.

The gravitational field equations are obtained by varying $L_G$ with respect to $b^i$ and $\om^{ij}$. They can be written in a compact form as
\bsubeq\lab{2.1}
\bea
    \d b^i:&&\quad \nab H_i+E_i=0\, ,                                \\
\d\om^{ij}:&&\quad \nab H_{ij}+E_{ij}=0\, ,
\eea
\esubeq
were $H_i:=\pd L_G/\pd T^i$ and $H_{ij}:=\pd L_G/\pd R^{ij}$ (2-forms) are the covariant momenta, and $E_i:=\pd L_G/\pd b^i$ and $E_{ij}:=\pd L_G/\pd\om^{ij}$ (3-forms) are the corresponding energy-momentum and spin currents, respectively.

The content of the field equations \eq{2.1} depends on the structure of the Lagrangian. In the present work, the PG Lagrangian is assumed to have the most general form, with all possible parity even and parity odd terms,
\bsubeq\lab{2.2}
\bea
&&L_G=L^+_G+L^-_G\,,                                                 \\
&&L_G^+:=-\hd\big(a_0R+2\L\big)+T^i\sum_{n=1}^3\hd\big(a_n\ir{n}T_i\big)
         +\frac{1}{2}R^{ij}\sum_{n=1}^6\hd\big(b_n\ir{n}R_{ij}\big)\,, \\
&&L^-_G:=-\aa_0\hd X+T^i\sum_{n=1}^3\aa_n\ir{n}T_i
             +\frac{1}{2}R^{ij}\sum_{n=1}^6\bb_n\ir{n}R_{ij}\,.
\eea
\esubeq
Here, $(a_n,b_n,\L_0)$ and $(\aa_n,\bb_n)$ are the coupling constant, $R$ and $X$ are the scalar and pseudoscalar curvatures, $\hd R=\hd(b_ib_j)R^{ij}$ and $\hd X=b_ib_j R^{ij}$,  and $\ir{n}T^i$ and $\ir{n}R^{ij}$ are the irreducible components of the torsion and the curvature, respectively; see Ref. \cite{x6}. Because some terms in $L^-_G$ are the same, the corresponding coupling constants are not independent; in particular, one can choose $\aa_2=\aa_3$, $\bb_2=\bb_4$ and $\bb_3=\bb_6$, see \cite{x16,x17}. Further freedom in the choice of parameters follows from the existence of three topological invariants \cite{x13}.

With the above form of $L_G$, the explicit expressions for the covariant momenta read
\bsubeq\lab{2.3}
\bea
&&H_i=2 \sum_{n=1}^3\big[\hd\big(a_n\ir{n}T_i\big)+\aa_n T^i\big]\,, \nn\\
&&H_{ij}=-2a_0\hd(b_ib_j)-2\aa_0(b_ib_j)+H'_{ij}\,,
\eea
where
\be
H'_{ij}:=2\sum_{n=1}^6\big[\hd\big(b_n\ir{n}R_{ij}\big)+\bb_n\ir{n}R_{ij}\big]\,.
\ee
\esubeq
They play a crucial role not only in the structure of the field equations, but also, as we shall see, in the analysis of the conserved charges and entropy.

Following the ideas of Regge and Teitelboim  \cite{x18}, asymptotic charges can be introduced as certain boundary terms $\G$ associated to the naive canonical gauge generator $G$, which is weakly vanishing. Namely, if $G$ is not regular (differentiable), it can be improved by adding a suitable surface term $\G$, $\tG:=G+\G$, such that
\be
\d\tG=\d G+\d\G=\text{regular}\,.                                   \lab{2.4}
\ee
In Ref. \cite{x6}, this construction, combined with Wald's identification of entropy as the Noether charge on horizon \cite{x7}, is used to propose a \emph{unified approach} to both the asymptotic charges and black hole entropy in PG.

Next, consider a stationary black hole, such that its spatial section $\S$ has a boundary with two components, one at infinity and the other at horizon, $\pd\S=S_\infty\cup S_H$. The corresponding boundary integral has two parts, $\G=\G_\infty-\G_H$ (the minus sign reflects a different orientation of $S_H$), which are determined by the variational equations
\bsubeq\lab{2.5}
\bea
&&\d\G_\infty=\oint_{S_\infty}\d B(\xi)\,,\qquad
       \d\G_H=\oint_{S_H} \d B(\xi)\,,                               \\
&&\d B(\xi):=(\xi\inn b^{i})\d H_i+\d b^i(\xi\inn H_i)
   +\frac{1}{2}(\xi\inn\om^{ij})\d H_{ij}
   +\frac{1}{2}\d\om^{ij}(\xi\inn\d H_{ij})\, .
\eea
\esubeq
Here, $\xi$ is the Killing vector with values $\pd_t$ and $\pd_\vphi$ on $S_\infty$, and a linear combination thereof on $S_H$, such that $\xi^2=0$.
Moreover, a consistent interpretation of these equations is based on the following simple rules:
\bitem
\item[(r1)] On the boundary $S_\infty$, the variation $\d$ acts on the parameters of a black hole solution, but not on the parameters of the background configuration.\vsm
\item[(r2)] On $S_H$, the variation $\d$ must keep surface gravity constant.
\eitem
The boundary conditions must be chosen so as to ensure the solutions for $\G_\infty$ and $\G_H$ to exist and be finite ($\d$-integrability).
When these requirements are satisfied, $\G_\infty$ and $\G_H$ are interpreted as the asymptotic charges and entropy, respectively, of a stationary black hole.

Note that each covariant momentum is given as a sum of parity even and parity odd terms. This allows us to make the corresponding decomposition for the thermodynamic charges \eq{2.5} and simplify their calculation.

According to the variational equations \eq{2.5}, the boundary terms $\d\G_\infty$ and $\d\G_H$ are a priori independent quantities. However, since $\d\G=\d\G_\infty-\d\G_H$ is introduced to ensure the regularity of the canonical gauge generator $G$, see Eq. \eq{2.4}, one can conclude that if $G$ is regular then $\d\G=0$ by construction. Since the inverse statement is also true ($\d\G=0$ implies $G$ is regular), it follows that
\be
G \text{~~is~regular}\quad \Lra\quad \d\G\equiv \d\G_\infty-\d\G_H=0\, .
\ee
The statement $\d\G_\infty=\d\G_H$ is nothing but the \emph{first law} of black hole thermodynamics.

In the previous paper \cite{x10}, we studied the asymptotic charges and entropy of the Kerr-AdS black holes with torsion, found by Baekler et al. \cite{x19} in the parity even sector of PG. In the present work, we extend these considerations to the general PG Lagrangian \eq{2.2}. The corresponding Kerr-AdS black holes were constructed recently by Obukhov \cite{x16}.

\section{Kerr-AdS solutions in PG}\label{sec3}
\setcounter{equation}{0}

We find it convenient to introduce the symbols \pgp\ and \pgm\ referring to the parity sectors of PG, defined by the Lagrangians $L_G^+$ and $L_G^-$, respectively.

\subsection{Geometry}

The general Kerr-AdS solution in PG \cite{x16} and its PG$^+$ counterpart \kadsp\ \cite{x19,x10} are defined by the same tetrad field; in Boyer-Lindquist coordinates, it has the form
\bsubeq\lab{3.1}
\bea
&&b^0=N\Big(dt+\frac{a}{\a}\sin^2\th\,d\vphi\Big)\,,\qquad
  b^1=\frac{dr}{N}\,,                                                \nn\\
&&b^2=Pd\th\, ,\qquad
  b^3=\frac{\sin\th}{P}\Big[a\,dt+\frac{(r^2+a^2)}{\a}d\vphi\Big]\,,
\eea
where
\bea
&&N=\sqrt{\D/\r^2}\, ,\qquad \r^2=r^2+a^2\cos^2\th\,,                \nn\\
&&\D=(r^2+a^2)(1+\l r^2)-2mr\,,\qquad \a=1-\l a^2\,,                 \nn\\
&&P=\sqrt{\r^2/f}\,, \qquad  f=1-\l a^2\cos^2\th\, .
\eea
\esubeq
Here, $m$ and $a$ are the parameters of the solution,  $0\le\th<\pi$, $0\le\vphi<2\pi$ and $a_0\l=-\L/3$.

The metric $ds^2=\eta_{ij}b^i\otimes b^j$, which is stationary and axially symmetric, admits the Killing vectors $\pd_t$ and $\pd_\vphi$. The metric characteristics of Kerr-AdS black holes remain the same as for \kadsp. In particular, this holds, respectively, for the location of the outer horizon  $r=r_+$, the horizon area $A_H$, the angular velocity $\om_+$ and the surface gravity $\k$,
\bsubeq
\bea
&&\D(r_+)\equiv(r_+^2+a^2)(1+\l r_+^2)-2mr_+=0\,,                       \\
&&A_H=\int_{r_+}b^2b^3=4\pi\frac{r_+^2+a^2}{\a}\,,                      \\
&&\om_+=\left.\frac{g_{t\vphi}}{g_{\vphi\vphi}}\right|_{r_+}
                                  =\frac{a\a}{r_+^2+a^2}\,,             \\
&&\k=\frac{[\pd\D]_{r_+}}{2(r_+^2+a^2)}\,.
\eea
\esubeq

By construction, the \kads\ and \kadsp\ black holes also have the same torsion,
\bsubeq\lab{3.3}
\bea
&&T^0:=\frac{1}{N}\big(-V_1b^0b^1-2V_4b^2b^3\big)
          +\frac{1}{N^2}b^-\big(V_2b^2+V_3b^3\big)=:T^1\,,                \nn\\
&&T^2:=\frac{1}{N}b^-\big(V_5b^2+V_4b^3\big)\,,                      \nn\\
&&T^3:=\frac{1}{N}b^-\big(-V_4b^2+V_5b^3\big)\,,                   \lab{3.3a}
\eea
where $b^-:=b^0-b^1$ and the torsion functions $V_n$ are
\bea
&&V_1=\frac{m}{\r^4}(r^2-a^2\cos^2\th)\, ,\qquad
  V_2=-\frac{m}{\r^4 P}ra^2\sin\th\cos\th\, ,                        \nn\\
&&V_3=\frac{m}{\r^4 P}r^2a\sin\th\, ,\qquad
  V_4=\frac{m}{\r^4}ra\cos\th\,,\qquad V_5=\frac{m}{\r^4}r^2\,.
\eea
\esubeq
The third irreducible part of $T^i$ vanishes.

For a given torsion, one can introduce the RC connection 1-form  by
\be
\om^{ij}:=\tom^{ij}+K^{ij}\,,
\ee
where $\tom^{ij}$ is the Riemannian connection and $K_{ij}$ the contortion 1-form,
\be
K^{ij}=\frac{1}{2}\Big[h^i\inn T^j-h_j\inn T^i
                     -\Big(h^i\inn\big(h^j\inn T^k\big)\Big)b_k\Big]\, .
\ee
Clearly, the connection is the same for both Kerr-AdS and \kadsp.

The corresponding RC curvature $R^{ij}=d\om^{ij}+\om^i{}_k\om^{kj}$ has only two nonvanishing irreducible parts;  with $A=(0,1)$ and $c=(2,3)$, they are
\be
\ir{6}R^{ij}=\l b^ib^j\, ,\qquad
             \ir{4}R^{Ac}=\frac{\l m r}{\D}b^-b^c\,.
\ee
The quadratic invariants are regular,
\be
R^{ij}\hd R_{ij}=12\l^2\heps\,,\qquad T^i\hd T_i=0\,.
\ee

\subsection{Dynamics}

The Lagrangian parameters of \kads\ solutions are restricted by the conditions \cite{x16}
\bsubeq\lab{3.8}
\bea
&&2a_1+a_2=0\,,\quad a_0-a_1-\l(b_4+b_6)=0\,,\quad a_0\l=-\L/3\,,\lab{3.8a}\\
&&\aa_2-\aa_1=0\, ,\qquad \aa_0-\aa_1+\l(\bb_4-\bb_6)=0\,.       \lab{3.8b}
\eea
\esubeq
imposed by the field equations.

Although the dynamical variables $(b^i,\om^{ij})$ have  the same form for both Kerr-AdS and \kadsp, the corresponding Lagrangians $L_G$ and $L_G^+$ are different. To clarify dynamical aspects of this difference,
recall that each of the covariant momenta $H_i$ and $H_{ij}$, defined in Eqs. \eq{2.3}, contains two terms, one coming from $L^+_G$ and the other from $L^-_G$. As a consequence, the formulas \eq{2.5} for the boundary terms imply:
\bitem
\item[(i)] The thermodynamic charges of Kerr-AdS black holes can be obtained by summing up the contributions stemming from $L_G^+$ and $L_G^-$.
\eitem
Geometrically, the first term is associated to \kadsp\ black holes\footnote{A comment on the geometric aspects of the the second one is given in the last section.}. Since the thermodynamic charges for \kadsp\ are already known, see Ref. \cite{x10}, it remains only to calculate the contributions stemming from $L_G^-$. The related covariant momenta are determined by the effective form of the parity odd Lagrangian,
\be
L^-_G=-\aa_0\hd X+T^i(\aa_1\ir{1}T_i+\aa_2\ir{2}T_i)
        +\frac{1}{2}R^{ij}(\bb_4\ir{4}R_{ij}+\bb_6\ir{6}R_{ij})\, ,   \lab{3.9}
\ee
containing only the nonvanishing irreducible parts of the field strengths. Thus,
\bea
\bH_i&=&2\,(\aa_1\ir{1}T_i+\aa_2\ir{2}T_i)=2\aa_1 T_i\, ,               \nn\\
\bH_{ij}&=&-2(\aa_0-\l \bb_6) b_ib_j+2\bb_4\ir{4}R_{ij}\,.
\eea
The dynamical difference between Kerr-AdS and \kadsp, including the values of their thermodynamic charges, is hidden just in the above expressions.

\section{Asymptotic charges}\label{sec4}
\setcounter{equation}{0}

Asymptotic conditions determine the behaviour of dynamical variables on the boundary $S_\infty$ where the asymptotic charges are calculated. Hence, a precise definition of the asymptotic charges requires to have a definite choice of the background configuration. For a Kerr-AdS black hole, the background is defined by $m=0$ and interpreted as the standard AdS spacetime, with vanishing torsion and constant curvature. Note, however, that the AdS metric in the Boyer-Lindquist coordinates depends on the parameter $a$, which complicates the variational procedure introduced in Sec. \ref{sec2}. Namely, according to the rule (r1), the variation over parameter $a$ appearing in the AdS configuration should be avoided. How to recognise those unneeded $\d a$ terms? Technically, this can be taken care of by an improved version of the rule (r1), specifically designed for \kads\ black holes, which is as follows:
\bitem
\item[(r1$^\prime$)] In the variational equation \eq{2.5} for $\G_\infty$, the variation $\d$ is first applied to all the parameters $(m,a)$ appearing in $B(\xi)$. Then, those $\d a$ terms that survive the limit $m=0$ have to be disregarded, as they stem from the variation of the AdS background.
\eitem

By a careful analysis of the asymptotic states, Henneaux and Teitelboim \cite{x20} concluded that the \kads\ metric in Boyer-Lindquist coordinates does not obey the asymptotic conditions compatible with the standard AdS background; see also Carter \cite{x21}. The problem was resolved using a suitable coordinate transformation which brings the metric to a manifestly asymptotically AdS form. In our approach, based on the variational equations \eq{2.5}, the inadequacy of the Boyer-Lindquist coordinates becomes visible through the lack of $\d$-integrability. As we argued in \cite{x9,x10}, the problem can be solved by going over to the ``untwisted" coordinates
\bsubeq\lab{4.1}
\be
T=t\,,\qquad \phi=\vphi-\l a t\,.                                    \lab{4.1a}
\ee
Namely, if
$\d E_t:=\d\G_\infty(\pd_t)$ and $\d E_\vphi:=\d\G_\infty(\pd_\vphi)$ are taken as the naive expressions for the asymptotic charges, then their $(T,\phi)$ transforms
\be
\d E_T=\d E_t+\l a\d E_\vphi\,,\qquad \d E_\phi=\d E_\vphi\,,        \lab{4.1b}
\ee
\esubeq
become $\d$-integrable.

In further analysis, we shall focus on the unknown thermodynamic charges associated to the parity odd sector \pgm. Using the rule (r1$^\prime$), we will first calculate the naive expressions for $\d E_\vphi$ and $\d E_t$, whereupon \eq{4.1b} will produce the final, $\d$-integrable results.

\subsection{Angular momentum}

Consider the expression $\bE_\vphi$, defined by the variational equation $\d\bE_\vphi:=\d\G_\infty(\pd_\vphi)$, where the covariant momenta are restricted to the \pgm\ sector. To calculate $\d\bE_\vphi$, we rewrite it in the form $\d \bE_\vphi=\d\bE_{\vphi 1}+\d\bE_{\vphi 2}$, where
\bea
&&\d\bE_{\vphi 1}:=\frac{1}{2}\om^{ij}{}_\vphi\d\bH_{ij}
                 +\frac{1}{2}\d\om^{ij}\bH_{ij\vphi}\, ,             \nn\\
&&\d\bE_{\vphi 2}:=b^i{}_\vphi\d\bH_i+ \d b^i\bH_{i\vphi}\,,
\eea
and the integration over $S_\infty$ is implicitly understood.
Using the relations \eq{A.1}, one finds that the nontrivial content of $\d \bE_{\vphi 1}$ is given by
\be
\d\bE_{\vphi 1}
  =\d\big(\om^{02}{}_\vphi\bH_{02\th\vphi}+\om^{12}{}_\vphi\d\bH_{12\th\vphi}
            +\om^{23}{}_\vphi\d\bH_{23\th\vphi} \big)d\th d\vphi\,.  \nn
\ee
For large $r$, $\d\bE_{\vphi 1}$ is quadratically divergent,
\be
\d\bE_{\vphi 1}=\a_2\l r^2+\a_0+O_1\, .                           \lab{4.3}
\ee
Since the coefficients $\a_2$ and $\a_0$ contain only \emph{$m$-independent} $\d a$ terms, the rule (r1$^\prime$) implies that their contribution should be disregarded. Hence, the integration over $S_\infty$ yields effectively $\d\bE_{\vphi 1}=0$. Similar analysis of
\be
\d\bE_{\vphi 2}=\d(b^0{}_\vphi\bH_{0\th\vphi}
                       + b^3{}_\vphi\bH_{3\th\vphi})d\th d\vphi      \nn
\ee
yields $\d\bE_{\vphi 2}=0$. Transition to the new coordinates $(T,\phi)$ implies $\d\bE_\phi=\d\bE_\vphi$. Hence:
\bitem
\item[(ii)] The variation of angular momentum associated to the \pgm\ sector vanishes,
\be
\d\bE_\phi=\d\bE_{\vphi 1}+\d\bE_{\vphi 2} =0\,.
\ee
\eitem

\subsection{Energy}

Following the procedure used for \kadsp\ black holes \cite{x9,x10}, energy stemming from the \pgm\ sector is calculated in two steps. First,  consider the naive expression $\bE_t:=\d\G_\infty(\pd_t)$, where the covariant momenta are restricted to the \pgm\ sector and represented it as the sum of two terms,
\bea
&&\d\bE_{t1}=\frac{1}{2}\om^{ij}{}_t\d\bH_{ij}
                        +\frac{1}{2}\d\om^{ij}\bH_{ijt}\, ,          \nn\\
&&\d\bE_{t2}=b^i{}_t\d\bH_{i}+\d b^i\bH_{it}\,.
\eea
Then, the relations \eq{A.2} allow one to reduce the form of $\d\bE_{t1}$ as
\be
\d\bE_{t1}=\big(\om^{23}{}_t\d\bH_{23\th\vphi}
            +\d\om^{03}{}_\th\bH_{03t\vphi}+\d\om^{13}{}_\th\bH_{13t\vphi}
               -\d\om^{02}{}_\vphi\bH_{02t\th}
               -\d\om^{23}{}_\vphi\bH_{23t\th}\big)d\th d\vphi\,.    \nn
\ee
For large $r$, $\d\bE_{t1}$ is found to be
\be
\d\bE_{t1}=\b_2 \l r^2+\b_0+O_1\,,
\ee
where $\b_2$ and $\b_0$ are \emph{$m$ independent} and proportional to $\d a$; thus, the integration over $S_\infty$ yields $\d\bE_{t1}=0$. In a similar manner, the relation
\be
\d\bE_{t2}=\big(b^0{}_t\d\bH_0+b^3{}_t\d\bH_3+\d b^0\bH_{0t}
      +\d b^2\bH_{2t}+\d b^3\bH_{3t}\big)d\th d\vphi                 \nn
\ee
implies that $\d\bE_{t2}$ also vanishes. Hence,
$\d\bE_t\equiv\d\bE_{t1}+\d\bE_{t2}=0$.

In the second step, after going over to the $(T,\phi)$ coordinates, one finds:
\bitem
\item[(iii)] The variation of energy associated to the \pgm\ sector also vanishes,
\be
\d\bE_T=\d\bE_t+\l a\d\bE_\vphi=0\,.
\ee
\eitem

\section{Entropy}\label{sec5}
\setcounter{equation}{0}

Consider the variational equation for $\d\G_H$ restricted to the \pgm\ sector, where
\bsubeq
\be
\xi:=\pd_T-\Om_+\pd_\vphi\,,\qquad
     \Om_+:=\om_++\l a=\frac{a(1+\l r_+^2)}{r_+^2+a^2}\,,
\ee
and $\Om_+$ is the angular velocity in the new coordinates $(T,\phi)$. Moreover, we use the notation
\bea
&&\bA_0:=\aa_0-\l\bb_6\,,                                           \nn\\
&&Y^A{}_\xi:=\xi\inn Y^A\,,\quad\text{where}\quad
  Y^A=(b^i,\om^{ij},H_i,H_{ij})\,.
\eea
\esubeq
For convenience, the expression $\d\G_H$ is divided into two parts, $\d\G_1$ and $\d\G_2$, and in further calculations, we rely on Eq. \eq{A.3}.

\subsubsection*{\mb{\d\G_1=\frac{1}{2}\om^{ij}{}_\xi\d H_{ij}
                            +\frac{1}{2}\d\om^{ij}H_{ij\xi}}}

The only nontrivial contributions to the first term in $\d\G_1$ are
\bsubeq\lab{5.2}
\bea
&&\om^{23}{}_\xi\d H_{23\th\vphi}=K^{23}{}_\xi \d H_{23\th\vphi}
  =2\bA_0\cdot
   \frac{2amr_+}{(r_+^2+a^2)\r_+^2}\d\Big(\frac{r_+^2+a^2}{\a}\Big)
                                               \sin\th\cos\th\,,\lab{5.2a}\\
&&\om^{02}{}_\xi\d H_{02\th\vphi}+\om^{12}{}_\xi\d H_{12\th\vphi}
  =\tom^{02}{}_\xi\d H_{02\th\vphi}
           +K^{02}{}_\xi\d(H_{02\th\vphi}+H_{12\th\vphi})            \nn\\
&&=\left[2\l\bb_4\frac{Na^2}{P(r_+^2+a^2)}
    \d\Big(\frac{mr_+}{N\r_+^2}P\frac{a}{\a}\Big)
     -2\bA_0\frac{a^2mr_+}{NP(r_+^2+a^2)\r_+^2}\d\Big(PN\frac{a}{\a}\Big)
                           \right]\sin^3\th\cos\th\,.\qquad
                                                                \lab{5.2b}
\eea
\esubeq
The second term in $\d\G_1$ is determined by
\bsubeq\lab{5.3}
\bea
&&\d\om^{03}H_{03\xi}+\om^{13}\d H_{13\xi}
  =\d\tom^{03}{}_\th H_{03\xi\vphi}
     +\d K^{03}{}_\th(H_{03\xi\vphi}+H_{13\xi\vphi})                 \nn\\
&&=\left[-2\l\bb_4\d\Big(\frac{aNP}{\r_+^2}\Big)\frac{mr_+}{NP\a}
  +2\bA_0\d\Big(\frac{amPr_+}{N\r_+^4}\Big)
                    \frac{N\r_+^2}{P\a}\right]\sin\th\cos\th\,. \lab{5.3a}
\eea
\bea
&&\d\om^{02}H_{02\xi}+\d\om^{12} H_{12\xi}
  =-\d\tom^{02}{}_\vphi H_{02\xi\th}
     -\d K^{02}{}_\vphi(H_{02\xi\th}+H_{12\xi\th})                   \nn\\
&&=\left[-2\l\bb_4\d\Big(\frac{aN}{P\a}\Big)\frac{mr_+P}{N(r_+^2+a^2)}
  +2\bA_0\d\Big(\frac{amr_+}{NP\a\r_+^2}\Big)
           \frac{NP\r_+^2}{r_+^2+a^2}\right]\sin\th\cos\th\,.   \lab{5.3b}
\eea
\esubeq

\subsubsection*{\mb{\d\G_2=b^i{}_\xi\d H_i+\d b^iH_{i\xi}}}

The analysis of the nontrivial content of $\d\G_2$ yields
\bsubeq\lab{5.4}
\bea
&&b^0{}_\xi\d H_0=b^0{}_\xi\d H_{0\th\vphi}
 =-2\aa_1\frac{N\r_+^2}{r_+^2+a^2}\cdot
   \d\Big[\frac{amr_+}{N\a\r_+^4}(r_+^2+a^2+\r_+^2)\Big]\sin\th\cos\th,
                                                           \lab{5.4a}\qquad\\
&&\d b^0 H_{0\xi}=-\d b^0{}_\vphi H_{0\xi\th}
  =2\aa_1\d\Big(\frac{Na}{\a}\Big)\cdot
            \frac{a^2mr_+}{N(r_+^2+a^2)\r_+^2}\sin^3\th\cos\th\,,
                                                                \lab{5.4b}\\
&&\d b^2 H_{2\xi}= \d P H_{2\xi\vphi}
  =-2\aa_1\d P\cdot\frac{amr_+}{P\a\r_+^2}\sin\th\cos\th\,,     \lab{5.4c}\\
&&\d b^3 H_{3\xi}=-\d b^3{}_\vphi H_{3\xi\th}
  =-2\aa_1\d\Big(\frac{r_+^2+a^2}{P\a}\Big)
       \cdot\frac{amr_+P}{(r_+^2+a^2)\r_+^2}\sin\th\cos\th\,.        \lab{5.4d}
\eea
\esubeq

\subsubsection*{Calculations and the result}

In order to obtain the entropy associated to the \pgm\ sector, one could now apply the systematic procedure formulated in Ref. \cite{x10} to calculate $\d\G_H\equiv \d\G_1+\d\G_2$.  However, the procedure can be enormously shortened by noting that each term in Eqs. \eq{5.2}, \eq{5.3} and \eq{5.4}, is given as an integral of the form
\be
I=\int_0^{\pi} d\th f(\cos^2\th)\cos\th\sin\th\,.
\ee
Then, the change of variables $x=\cos\th$ implies $I=0$, and consequently:
\bitem
\item[(iv)] The variation of entropy associated to the \pgm\ sector vanishes,
\be
\d\G_H\equiv T\d\bS=0\,.
\ee
\eitem

\section{Concluding remarks}\label{sec6}

In the present paper, we performed a Hamiltonian analysis of the thermodynamic
charges for \kads\ black holes in the general PG, with both even and odd parity modes.

Our methodology is compactly formulated by the variational equations \eq{2.5}, accompanied by the basic set of two rules, (r1)  and (r2), for the variation $\d$. When the background configuration is an AdS spacetime, the validity of the rule (r1) is ensured  by an additional instruction on how the variation of the background should be avoided, see (r1$^\prime$) in Sec. \ref{sec4}. These rules are a variational counterpart of the asymptotic conditions used in Refs. \cite{x20}, as well as in \cite{x5}$_2$, in their analyses of \kads\ spacetimes.

\kads\ solutions in PG can be understood as a superposition of two contributions, associated to the \pgp\ and \pgm\ sectors of PG. The thermodynamic charges originating from \pgm\ are found to be vanishing. Thus:
\bitem
\item[(v)] Asymptotic charges and entropy of \kads\ black holes in PG \cite{x16} coincide with the corresponding expressions for \kadsp\ black holes, found earlier in \pgp\ \cite{x10}. With $T:=\k/2\pi$, we have
\be
E_T=16\pi a_1\frac{m}{\a^2}\,,\qquad E_\phi=16\pi a_1\frac{ma}{\a^2}\,,
\qquad S=16\pi a_1\frac{A_H}{4}\,.
\ee
\eitem

Using the Pontryagin and Nieh-Yan topological invariants, the effective form of $L_G^-$ in \eq{3.9} can be reduced to just two terms, $\hd X$ and $R^{ij}\,\ir{6}R_{ij}\sim\hd X$. Thus, in spite of the fact that the thermodynamic charges in the parity odd sector vanish, the sector itself is not dynamically trivial---it is essentially equivalent to the Cartan-Holst term $\hd X$.

\kads\ solutions cannot be consistently reduced to the pure parity odd sector \pgm. Namely, for $a_0,\L=0$, the parameter $\l$ would remain undetermined (since $3a_0\l=\L$), which would be a degenerate situation. However, they can be consistently restricted to the \pgm\ sector extended by the nonvanishing $(a_0,\L)$. Although the resulting solution (\kadsm)$^\prime$ is interesting in its own right, the \kads\ spacetime is not a proper superposition of \kadsp\ and (\kadsm)$^\prime$, as their parameters overlap.

\section*{Acknowledgments}

We would like to thank Yuri Obukhov for a critical reading of the manuscript. This work was partially supported by the Ministry of Education,
Science and Technological development of the Republic of Serbia.

\appendix
\section{Useful formulas}\label{appA}
\setcounter{equation}{0}

In this appendix, we present technical formulas which are used in deriving the expressions for the asymptotic charges and entropy, in sections \ref{sec4} and \ref{sec5}.
\bea
&&H^{01}{}_{\th\vphi}=H^{03}{}_{\th\vphi}=H^{13}{}_{\th\vphi}=0\,,   \nn\\
&&b^1{}_\vphi=b^2{}_\vphi=0\,,\qquad
 \tom^{03}{}_\vphi=\tom^{12}{}_\vphi=0\,,                          \lab{A.1}
\eea
\bea
&&H^{01}{}_{t\vphi}=H^{02}{}_{t\vphi}=
  H^{12}{}_{t\vphi}=H^{23}{}_{t\vphi}=0\, ,                          \nn\\
&&H^{01}{}_{t\th}=H^{03}{}_{t\th}=H^{13}{}_{t\th}=0\, ,              \nn\\
&&b^1{}_t=b^2{}_t=0\,,\qquad \om^{03}{}_t=\om^{12}{}_t=0\,,          \nn\\ &&\tom^{01}{}_\th=\tom^{02}{}_\th=\tom^{13}{}_\th=\tom^{23}{}_\th=0\,.\lab{A.2}
\eea
\bea
&&b^0{}_\xi=N\frac{\r_+^2}{r_+^2+a^2}\qquad
  b^a{}_\xi\big|_{r_+}=0\quad a=1,2,3;                               \nn\\
&&H^{01}{}_\xi=H^{23}{}_\xi=0\,,\quad
  H^{03}{}_{\xi\th}=H^{13}{}_{\xi\th}=0\,,\quad H^{02}{}_{\xi\vphi}=H^{12}{}_{\xi\vphi}=0\,,                       \nn\\
&&\tom^{03}{}_\xi=\tom^{12}{}_\xi=\tom^{23}{}_\xi=0\,,
  \qquad \tom^{23}{}_\xi= O(N^2)\,,                                  \nn\\
&&\tom^{12}{}_\vphi=0\,,\quad \tom^{13}{}_\th=0\,,\quad K^{01}{}_\th=K^{23}{}_\th=0\,.                                    \lab{A.3}
\eea

\end{document}